\documentclass[preprint,aps, showpacs]{revtex4}
\usepackage[LY1]{fontenc}       
\usepackage[LY1,mtbold]{mathtime}   
\usepackage{times}          
\usepackage{graphicx}
\usepackage{bm}
\usepackage{amsfonts}
\usepackage{amssymb}

\newcommand{\abs}[1]{\left\vert#1\right\vert}

\newcommand{\ket}[1]{\left\vert #1\right\rangle}

\newcommand{\brkt}[2]{\left\langle #1 \vert #2\right\rangle}

\newcommand{\braket}[3]{\left\langle #1 \right\vert #2\left\vert #3\right\rangle}

\begin{document}


\title{{\large Gamma N Delta Form Factors and Wigner Rotations}}
\author{Milton Dean Slaughter}
\address{Department of Physics, Florida International University, Miami, Florida 33199}
 \email{E-Mail: slaughtm@FIU.Edu}

\begin{abstract}
For more than 50 years the $\Delta
N\gamma$ form factors have been studied experimentally, theoretically, and phenomenologically.  Although there has been substantial progress in understanding their behavior, there remains much work to be done.  A major tool used in many investigations is the Jones-Scadron $\Delta$ rest frame parametrization of the three $\Delta
N\gamma$ form factors.  We point out that many studies utilizing this parametrization  may not account for Wigner rotations and the consequent helicity mixing that ensues when the $\Delta$ is not at rest.

\end{abstract}


\maketitle













The study of the $\Delta N\gamma$ transition form factors $G_{M}^{\ast
}(q^{2})$, $G_{E}^{\ast }(q^{2})$, and $G_{C}^{\ast }(q^{2})$ associated with the
$\Delta(1232)$ nucleon resonance isobar has engendered much
experimental and theoretical research for many decades.  As the most
studied nucleon resonance---the $\Delta(1232)$ has proved to be
very difficult in the determination of its physical properties
vis-\'{a}-vis its relation to the
nucleon.  Although
similar to the nucleon in valence quark content, it has spin and
isospin of $3/2$ as opposed to $1/2$ for the nucleon and thus its interaction with other particles via form
factors is much more complex than that of the nucleon.  In addition, the
$\Delta(1232)$ is unstable with a large width, making measurement of physical observables
 and
theoretical modeling much more difficult as well. Of special interest is the
$\Delta(1232)$--nucleon 4-vector electromagnetic current matrix
element in momentum space $\left\langle N\right| j^{\mu }(0)\left| \Delta
\right\rangle$ associated
with the process $\Delta \longleftrightarrow N+\gamma^{*}$
described covariantly by the form factors $G_{M}^{\ast }(q^{2})$,
$G_{E}^{\ast }(q^{2})$, and $G_{C}^{\ast }(q^{2})$, where $q^{2}$
is the photon 4-momentum transfer squared. This matrix element and
associated form factors is important in pion photoproduction and
electroproduction ({\it i.e.} $\pi N\rightarrow \Delta\rightarrow
\pi N\gamma$ or $\pi N$ and $\gamma N\rightarrow \Delta\rightarrow
\pi N$ or $\pi N\gamma$). In a world with unbroken
$SU_{F}(N)$ flavor symmetry, one
expects that $G_{E}^{\ast }(q^{2})=G_{C}^{\ast }(q^{2})=0$ and that $%
G_{M}^{\ast }(q^{2})$ would exhibit the same $q^{2}$ behavior as
does the Sachs nucleon form factor $G_{M}$ thus giving rise to
{\it pure} magnetic dipole $\Delta N\gamma$ transitions. Instead,
one finds that $G_{M}^{\ast }$ appears to
decrease faster as a function of $Q^{2}\equiv -q^{2}$ than does
$G_{M}$, the ratio $-G_{E}^{\ast }/G_{M}^{\ast }\neq 0$, the magnitude of $G_{E}^{\ast }(Q^{2})$ is small
when compared to $G_{M}^{\ast }(q^{2})$ near $Q^{2}\approx 0$, and that
$G_{M}^{\ast }$ possesses a complicated behavior as a function of
$Q^{2}$.

Probably the most widely used parametrization for the study of the $\Delta N\gamma$ transition form factors is that to Jones and Scadron~\cite{Jones:1972ky} (JS) followed by very closely related variations such as the helicity form factors of Devenish, Eisenschitz, and Korner~\cite{Devenish:1975jd} (DEK) or Bjorken and Walecka~\cite{Bjorken:1966ij}.  The JS parametrization is written explicitly in the the \emph{rest frame} of the $\Delta$ and introduces covariant couplings  $G_{M}^{\ast }(q^{2})$,
$G_{E}^{\ast }(q^{2})$, and $G_{C}^{\ast }(q^{2})$ analogous to the  familiar Sachs nucleon form factors  $G_{M}(q^{2})$ and $G_{E}(q^{2})$.  \emph{The fact that the JS parametrization--though covariant--utilizes in a very explicit fashion the rest frame of the $\Delta$ is the subject of this work as significant modifications may be manifested due to the existence of Wigner rotations} \cite{Wigner:1939cj} \emph{of purely geometric origin}.

Setting notation, we normalize physical states with
$\brkt{\vec{p'}}{\vec{p}}={(}2\pi {)^{3}}\;2p^{\;0}\;\delta^{3}%
(\vec{p^{\prime}}\,-\vec{p})$. Dirac spinors
are normalized by $\bar{u}(p)u(p)=2m$ . Our conventions for
Dirac matrices are $\left\{  \gamma^{\mu},\gamma^{\nu}\right\}
=2g^{\mu\nu}$ with $\gamma _{5}\equiv
i\gamma^{0}\gamma^{1}\gamma^{2}\gamma^{3}$, where $g^{\mu\nu}=$
Diag $(1,-1,-1,-1)$. The Ricci-Levi-Civita tensor is defined by
$\varepsilon _{0123}=-\varepsilon^{0123}=1=\varepsilon_{123}$.  As usual, we use natural units where $\hbar=c=1$.

In general, one may
write for the $\Delta  N \gamma $ transition amplitude---and incorporating the DEK prescription for removal of kinematic singularities at threshold and pseudo-threshold via an explicit $Q^{+}Q^{-}$ factor (defined below)---we have the
following JS-DEK \emph{expression where the} $\Delta$ \emph{is at rest}:

\begin{equation}
\left\langle N\vec{p},\lambda _{N}\right| j_{\mu }(0)\left| \Delta %
\vec{p^{*}},\lambda _{\Delta }\right\rangle =e\bar{u}_{N}(\vec{p},\lambda _{N})\left[
\Gamma _{\mu \beta }\right] u_{\Delta }^{\beta }(\vec{p^{*}},\lambda
_{\Delta })  \label{eqrest}
\end{equation}

\noindent where $e=\sqrt{4\pi\alpha}$, $\alpha$\, is the fine structure constant,
and $u_{\Delta }^{\beta }$ is a Rarita-Schwinger spinor.

\begin{equation}
\begin{array}{ccl}
\Gamma _{\mu \beta } & = &
{\displaystyle{3(m^{*}+m) \over 2m}}%
(G_{M}^{*}-3G_{E}^{*})(Q^{+}Q^{-})\left[im^{*}q_{\beta }\epsilon _{\mu
}(qp\gamma )\right]
\\
& + &
{\displaystyle{3(m^{*}+m) \over 2m}}%
(G_{M}^{*}+G_{E}^{*})(Q^{+}Q^{-})\left[im^{*}q_{\beta }\epsilon _{\mu
}(qp\gamma )-2\epsilon _{\beta \sigma
}(p^{*}p)\epsilon _{\mu }^{\;\sigma }(p^{*}p)\gamma
_{5}\right] \\
& + &  \label{eqrest2}
{\displaystyle{3(m^{*}+m) \over m}}%
G_{C}^{*}(Q^{+}Q^{-})q_{\beta }\left[ [p\cdot qq_{\mu }-q^{2}p_{\mu
}]\right]\gamma _{5}
\\
& = &
-h_{2}(Q^{+}Q^{-})\left[im^{*}q_{\beta }\epsilon _{\mu
}(qp\gamma )\right]
\\
& - &
h_{3}(Q^{+}Q^{-})\left[im^{*}q_{\beta }\epsilon _{\mu
}(qp\gamma )-2\epsilon _{\beta \sigma
}(p^{*}p)\epsilon _{\mu }^{\;\sigma }(p^{*}p)\gamma
_{5}\right] \\
& + &
h_{1}(Q^{+}Q^{-})q_{\beta }\left[ [p\cdot qq_{\mu }-q^{2}p_{\mu
}]\right]\gamma _{5},
\end{array}
\label{eqrest1}
\end{equation}
where
\begin{eqnarray}
h_{2} &=&-\frac{3(m^{*}+m)}{2m}(G_{M}^{*}+G_{E}^{*}),  \nonumber \\
h_{3} &=&-\frac{3(m^{*}+m)}{2m}(G_{M}^{*}-3G_{E}^{*}),  \nonumber \\
h_{1} &=&\frac{3(m^{*}+m)}{m}G_{C}^{*}\;.  \label{eq:ffh}
\end{eqnarray}
In Eqs. (\ref{eqrest}) and (\ref{eqrest1}), the electromagnetic current
is denoted by $j_{\mu }$ which transforms like a Lorentz 4-vector, $q\equiv p^{\ast }-p$, $p^{\ast }$ and
$p$ are the four-momenta of the $\Delta $ and nucleon
respectively with  $p^{\ast }\equiv (m^{*},\vec{0})$ and $p\equiv (p^{0},\vec{p})$.
$q^{2}$ is the invariant 4-momentum transfer, $m^{\ast }$ is the $\Delta  $ mass, $m$ is the nucleon
mass, and the magnitude of the three-momentum of the photon in
the $\Delta$ rest frame $\equiv |\vec{q_{c}}|$; \, $
Q^{+}Q^{-}=
4{m^{*}}^{2}\; |\vec{q_{c}}|^{2}$ with $%
Q^{\pm }\equiv (m^{\ast }\pm m)^{2}-q^{2}$; $\lambda _{N}$ and
$\lambda _{\Delta }$ are the helicities of the nucleon and the $\Delta$
respectively.

Note that the first (fourth $h_{2}$), second (fifth $h_{3}$), and third (sixth $h_{1}$) terms in Eq.~(\ref{eqrest2})
induce transverse $\frac{1}{2}$, transverse $\frac{3}{2}$, and
longitudinal (scalar) helicity transitions, respectively, in the rest frame
of the $\Delta$.  $h_{1}$, $h_{2}$, and $h_{3}$ are the DEK \emph{helicity} form factors
 and $G_{M}^{*}$, $G_{E}^{*}$, and $G_{C}^{*}$ induce
magnetic, electric, and coulombic (scalar) multipole transitions respectively.

The magnetic, electric, and scalar multipole transition moments given by $%
M_{1^{+}}(q^{2})$, $E_{1^{+}}(q^{2})$, and $S_{1^{+}}(q^{2})$ can
be written in terms of $G_{M}^{*}(q^{2})$, $G_{E}^{*}(q^{2})$, and
$G_{C}^{*}(q^{2})$:

\begin{eqnarray}
M_{1^{+}} &=&\alpha _{1}\;\sqrt{Q^{-}}\;G_{M}^{*},  \label{multipole} \\
E_{1^{+}} &=&\alpha _{2}\;\sqrt{Q^{-}}\;G_{E}^{*},  \nonumber \\
S_{1^{+}} &=&\alpha _{3}\;Q^{-}\sqrt{Q^{+}}\;G_{C}^{*},  \nonumber
\end{eqnarray}
where
$\alpha _{1}$, $\alpha _{2}=-\alpha _{1}$, and
$\alpha _{3}$\ are
dependent on the $\Delta $\ mass $m^{*}$ and width at resonance and other parameters governing the process $\Gamma(\Delta\rightarrow\pi N)$.

However, generally (theoretically, experimentally, and phenomenologically), one is confronted with the transition amplitude given by:

\begin{equation}
\left< N\vec{s},\lambda _{N}| j_{\mu }(0)|\Delta %
\vec{t},\lambda _{\Delta }\right>,   \label{eqgeneral}
\end{equation}

\noindent $s\equiv s^{\mu}=(s^{0},\vec{s})$, $t\equiv t^{\mu}=(t^{0},\vec{t})$
where the $\Delta$ is \emph{not at rest} and the nucleon and $\Delta$ 3-momenta \emph{are not necessarily collinear}.  \emph{Then one finds that matrix elements} Eq.~(\ref{eqrest}) and Eq.~(\ref{eqgeneral}) \emph{are related through a sequence of Lorentz transformations and consequent Wigner rotation angles} \cite{MartinSpearman:EPT,Perl:HEHP,Carruthers:SIPP,GibsonPollard:SPEPP,Gasiorowicz:EPT,Durand:1962zza}.

Let $U(\Lambda_{\chi})$ be an proper homogeneous Lorentz transformation operator which brings the $\Delta$ represented by the single-particle helicity state $\left| \Delta %
\vec{t},\lambda _{\Delta }\right\rangle$ to rest $\left| \Delta \vec{0},\lambda _{\Delta }\right\rangle$ and
$U(H(\overrightarrow{{\Lambda_{\chi} t}}))$ defines the transformation for the
helicity state $|N\vec{s},\lambda_{N}>=|Ns\theta\phi
\lambda_{N}>=U(R^{N}(\theta,\phi))|Ns\lambda_{N}>=U(R^{N}(\theta,\phi)Z_{N}(s))|N\lambda_{N}>\equiv U(H_{N}(\vec{s}))|N\lambda_{N}>$, thus $H_{N}(\vec{s})\equiv R^{N}(\theta,\phi)Z_{N}(s)$ is the homogeneous Lorentz transformation which defines the single-particle helicity state $|N\vec{s},\lambda_{N}>$---$R^{N}(\theta,\phi)$ is a rotation with Euler angles $\theta$ and $\phi$ and $Z_{N}(s)$ is a pure Lorentz boost such that $(m,\vec{0}){\stackrel{Z_{N}}{\rightarrow} } (s^{0},|\vec{s}|\;\hat{z})$.  Similarly, $H_{\Delta}(\vec{t})$ defines the state $|\Delta(\vec{t},\lambda_{\Delta})>=U(H_{\Delta}(\vec{t}))|\Delta\lambda_{\Delta}>$. $\ket{N\lambda_{N}}$ and $\ket{\Delta\lambda_{\Delta}}$ are rest frame states for the nucleon and $\Delta$ respectively.  Note that the quantization axis $\hat{z}$ is defined by $\vec{t}=\abs{\vec{t}}\hat{z}$ and that $p=\Lambda_{\chi}s$.

$\Lambda_{\chi}$ is a Lorentz transformation such that ${(\Lambda_{\chi})^{\mu}}_{\nu}\,r^{\nu}=r'^{\mu}=(r'^{0},\vec{r'})=
(r'^{0},\overrightarrow{\Lambda_{\chi}r})$, where $r$ and $r'$ are four-momenta $\Rightarrow$

\begin{eqnarray}
U(\Lambda_{\chi})\ket{N\vec{s},\lambda}&=&\sum_{{\lambda}'}\ket{N\overrightarrow{\Lambda_{\chi}s}\lambda'}
 \braket{N\overrightarrow{\Lambda_{\chi}s}\lambda'}{U(\Lambda_{\chi})}{N\vec{s},\lambda} \\
&=&\sum_{{\lambda}'}\ket{N\overrightarrow{\Lambda_{\chi}s}\lambda'}
\braket{N\lambda'}{U(H^{-1}_{N}(\overrightarrow{{\Lambda_{\chi} s}})\Lambda_{\chi}
H_{N}(\overrightarrow{{s}}))}{N,\lambda} \nonumber \\
&=&\sum_{{\lambda}'}\ket{N\overrightarrow{\Lambda_{\chi}s}\lambda'}
D_{{{\lambda'}},{\lambda}}^{(1/2)}(R_{W}^{N})  \nonumber
,
\end{eqnarray}

\noindent where $D_{{{\lambda'}},{\lambda}}^{(1/2)}(R_{W}^{N})=\braket{N\lambda'}{U(R_{W}^{N}(\Lambda_{\chi}s,s))}{N,\lambda}=
\braket{N\lambda'}{U(R_{W}^{N})}{N,\lambda}$, and the Wigner rotation which connects the rest frame nucleon states is given by
$R_{W}^{N}(\Lambda_{\chi}s,s)=
H^{-1}_{N}(\overrightarrow{{\Lambda_{\chi} s}})\Lambda_{\chi}
H_{N}(\overrightarrow{{s}})\equiv R_{W}^{N}$.  Analogously, for the $\Delta$, we have
$R_{W}^{\Delta}(\Lambda_{\chi}t,t)=
H^{-1}_{\Delta}(\overrightarrow{{\Lambda_{\chi} t}})\Lambda_{\chi}
H_{\Delta}(\overrightarrow{{t}})\equiv R_{W}^{\Delta}$.  Generally, $D_{{{\lambda'}},{\lambda}}^{(j)}(R_{W})$ is the Wigner rotation matrix for a particle of spin $j$ \cite{Wigner:1939cj,rose1957:etam}.

The helicity representation transformation law for $\Delta N \gamma$ matrix elements involving an operator $A$ (for clarity, we suppress any contravariant or covariant Lorentz indices here) which transforms under the Lorentz transformation $\Lambda_{\chi}$ like $U(\Lambda_{\chi})AU(\Lambda_{\chi})^{-1}=A_{\Lambda_{\chi}}$ is then given by \cite{Durand:1962zza}

\begin{equation}
\left< N\vec{s},\lambda _{N}|A|\Delta %
\vec{t},\lambda _{\Delta }\right>=\sum_{\lambda
{\lambda}'}D_{{{\lambda}^\prime},\lambda _{N}}^{(1/2)*}(R_{W}^{N})<N\overrightarrow{{\Lambda_{\chi} s}}{\lambda'}|A_{\Lambda_{\chi}}|\Delta \overrightarrow{{\Lambda_{\chi} t}}{\lambda}>
D_{{\lambda},{\lambda _{\Delta }}}^{(3/2)}(R_{W}^{\Delta}).
\label{eqtransform}
\end{equation}

For the case where $U(\Lambda_{\chi})$ is a pure homogeneous Lorentz transformation operator along the $-\hat{z}$ direction associated with a velocity $v_{\chi}$ and which brings the $\Delta$ represented by the single-particle helicity state $\left| \Delta (%
\vec{t},\lambda _{\Delta })\right\rangle$ to rest $| \Delta (\vec{0},\lambda _{\Delta })>$, \emph{i.e.} $\Lambda_{\chi}(\chi,\hat{z})\vec{t}=\vec{0}$ with $\vec{t}=(0,0,t_{z})$, then
Eq.~(\ref{eqtransform}) greatly simplifies since $H^{-1}_{\Delta}(\overrightarrow{{\Lambda_{\chi} t}})\Lambda_{\chi}
H_{\Delta}(\overrightarrow{{t}})=H^{-1}_{\Delta}(\vec{0})\Lambda_{\chi}
H_{\Delta}(\overrightarrow{{t}})=\mathbf{1} $ $\Rightarrow$
$D_{{\lambda},{{\lambda _{\Delta }}}}^{(3/2)}(R_{W}^{\Delta})=\delta_{\lambda,\lambda _{\Delta }}$ and we obtain

\begin{equation}
\left< N(\vec{s},\lambda _{N})|A|\Delta (%
\vec{t},\lambda _{\Delta })\right>=\sum_{{\lambda}'}D_{{{\lambda}^\prime},\lambda _{N}}^{(1/2)*}(R_{W}^{N})<N\overrightarrow{{\Lambda_{\chi} s}}{\lambda'}|A_{\Lambda_{\chi}}|\Delta \vec{0}{\lambda _{\Delta }}>.
\label{eqtransform1}
\end{equation}

Setting $A_{\Lambda_{\chi}}=\Lambda_{\chi}j^{\mu }{\Lambda_{\chi}^{-1}}$ one has

\begin{eqnarray}
\left< N\vec{s},\lambda _{N}|j^{\mu }|\Delta %
\vec{t},\lambda _{\Delta }\right>&=&\sum_{{\lambda}'}D_{{{\lambda}^\prime},\lambda _{N}}^{(1/2)*}(R_{W}^{N})<N\overrightarrow{{\Lambda_{\chi} s}}{\lambda'}|\Lambda_{\chi}j^{\mu }{\Lambda_{\chi}^{-1}}|\Delta \vec{0}{\lambda _{\Delta }}>  \label{eqtransform2}\\
&=&D_{-\frac{1}{2},\lambda _{N}}^{(1/2)*}(R_{W}^{N})<N\overrightarrow{{\Lambda_{\chi} s}}{-\frac{1}{2}}|\Lambda_{\chi}j^{\mu }{\Lambda_{\chi}^{-1}}|\Delta \vec{0}{\lambda _{\Delta }}> \nonumber \\
&&+
D_{\frac{1}{2},\lambda _{N}}^{(1/2)*}(R_{W}^{N})<N\overrightarrow{{\Lambda_{\chi} s}}{\frac{1}{2}}|\Lambda_{\chi}j^{\mu }{\Lambda_{\chi}^{-1}}|\Delta \vec{0}{\lambda _{\Delta }}> .  \nonumber
\end{eqnarray}

Eq.~(\ref{eqtransform2}) \emph{is the main result of this work and demonstrates explicitly the helicity mixing that occurs when the} JS $\Delta N\gamma$ \emph{form factor construction is used.  The Wigner rotation matrices are automatically brought into play and must be considered in most circumstances}. There are exceptions as will be made clearer below.

Without loss of generality, we consider only $\hat{x}-\hat{z}$ plane dynamics, where for instance, \emph{for the transverse components} $\mu=1$ or $\mu=2$, $\Lambda_{\chi}j^{\mu }{\Lambda_{\chi}^{-1}}=j^{\mu }$ .  The polar angles of $\vec{s}$ are given by $(\theta,\phi=0)$ and and the polar angles (all referred to the $\hat{z}$ axis) of $\vec{p}=\overrightarrow{\Lambda_{\chi}s}$ are given by $(\theta',\phi'=0)$.  Thus, the four-momentum vectors $s$ and $p$ are related by $p=\Lambda_{\chi}s$, whereas $p^{*}=(m^{*},\vec{0})=\Lambda_{\chi}t$.  We calculate the Wigner rotation angle with:
\begin{eqnarray}
{(\Lambda)^{\mu}}_{\nu}\,(\bar{\sigma},\bar{\theta})&=&
\left(
\begin{array}{llll}
 \cosh (\bar{\sigma} ) & 0 & 0 & \sinh
   (\bar{\sigma} ) \\
 \sin (\bar{\theta} ) \sinh (\bar{\sigma} ) &
   \cos (\bar{\theta} ) & 0 & \cosh (\bar{\sigma}
   ) \sin (\bar{\theta} ) \\
 0 & 0 & 1 & 0 \\
 \cos (\bar{\theta} ) \sinh (\bar{\sigma} ) &
   -\sin (\bar{\theta} ) & 0 & \cos (\bar{\theta}
   ) \cosh (\bar{\sigma} )
\end{array}
\right) \\
R_{W}(\omega)&=&
\left(
\begin{array}{llll}
 1 & 0 & 0 & 0 \\
 0 & \cos (\omega ) & 0 & \sin
   (\omega ) \\
 0 & 0 & 1 & 0 \\
 0 & -\sin (\omega ) & 0 & \cos
   (\omega )
\end{array}
\right)
\end{eqnarray}

Thus, we find that---$R_{W}^{N}(\Lambda_{\chi}s,s)=
\Lambda^{-1}_{N}(\sigma',\theta')\Lambda_{\chi}
\Lambda_{N}(\sigma,\theta)$---
\begin{eqnarray}
\tan(\omega_{N})&=&\frac{-\sin (\theta ) \sinh (\chi
   )}{\cosh (\chi ) \sinh (\sigma
   )-\cos (\theta ) \cosh (\sigma )
   \sinh (\chi )}=\frac{-\sin (\theta )\; m \abs{\vec{t}}
   }{t^{0} \abs{\vec{s}}-\cos (\theta ) \; s^{0}
   \abs{\vec{t}}}  \label{wigner1} \\
   &=&-\frac{\sin (\theta ) \abs{u_{\chi}]} \sqrt{1-v_s^2}}{\abs{v_s}-\cos (\theta )
   \abs{u_{\chi}]}}\;, \label{wigner2}
   \nonumber\\
\sin(\omega_{N})&=&\frac{-\sin (\theta ) \sinh (\chi
   )}{\sinh (\sigma')}=\frac{-\sin (\theta )\; m \abs{\vec{t}}}{m^{*} \abs{\vec{p}}}=\frac{-\sin (\theta ) \abs{u_{\chi }}{\sqrt{1-v_{p}^2}}}{\sqrt{{1-u_{\chi }^2}}\abs{v_{p}}}. \label{wigner3}
\end{eqnarray}

In Eq.~(\ref{wigner1}) and Eq.~(\ref{wigner3}), $\sinh (\chi)=\abs{u_{\chi }}({1-u_{\chi }^2})^{-1/2}$ where $u_{\chi}$ is the velocity parameter which specifies the Lorentz boost $\Lambda_{\chi}$, $v_s$ is the velocity of the nucleon in the frame where $\abs{\vec{s}}=m \sinh(\sigma)$ and $v_{p}$ is the velocity of the nucleon in the $\Delta$ rest frame. Thus, $\omega_{N}$ is independent of the nucleon mass and is a purely geometric phenomenon. Now $D_{{m},\; m'}^{(1/2)}(R_{W}^{N})=e^{-im\alpha}d_{{m},\; m'}^{(1/2)}(\omega_{N})e^{-im' \gamma}$, so choosing for conciseness $\alpha=\gamma=0$, then

\begin{eqnarray}
D_{{m},\; m'}^{(1/2)}(\omega_{N})=
\left(
\begin{array}{ll}
 \cos \left(\frac{\omega_{N} }{2}\right) & -\sin
   \left(\frac{\omega_{N} }{2}\right) \\
 \sin \left(\frac{\omega_{N} }{2}\right) & \cos
   \left(\frac{\omega_{N} }{2}\right)
\end{array}
\right).
\end{eqnarray}
\\
We give an example (transverse 1/2 transition, with $\mu=1-i2$, $\lambda_{\Delta}=1/2$, and $\lambda_{N}=-1/2$) using Eq.~(\ref{eqtransform2}):

\begin{eqnarray}
\left< N\vec{s},-1/2|j^{1-i2 }|\Delta %
\vec{t}=t_{z}\hat{z},+1/2\right>&=&\sum_{{\lambda}'}D_{{{\lambda}^\prime},-\frac{1}{2}}^{(-1/2)*}(R_{W}^{N})<N\overrightarrow{{\Lambda_{\chi} s}}{\lambda'}|j^{1-i2 }|\Delta \vec{0}{\frac{1}{2}}>  \label{eqtransform3}\\
&=&\cos \left(\frac{\omega_{N} }{2}\right)<N\overrightarrow{{\Lambda_{\chi} s}}{-\frac{1}{2}}|j^{1-i2 }|\Delta \vec{0}{\frac{1}{2}}> \nonumber \\
&&-
\sin \left(\frac{\omega_{N} }{2}\right)<N\overrightarrow{{\Lambda_{\chi} s}}{\frac{1}{2}}|j^{1-i2 }|\Delta \vec{0}{\frac{1}{2}}> .  \nonumber
\end{eqnarray}

We see that if $\overrightarrow{{\Lambda_{\chi} s}}$ is \emph{not collinear} with the $\hat{z}$ axis, $\sin \left(\frac{\omega_{N} }{2}\right)\neq 0$ and the matrix element $<N\overrightarrow{{\Lambda_{\chi} s}}{\frac{1}{2}}|j^{1-i2 }|\Delta \vec{0}{\frac{1}{2}}>$ is \emph{non-vanishing} as well.  There exist cases besides collinearity where the Wigner angle need not be calculated and the JS parametrization can be used without change:  A case example is when only the helicity-averaged quantity
$\frac{1}{2}$ $\sum_{\lambda _{N}\lambda _{\Delta }}\abs{\left< N\vec{s},\lambda _{N}|j^{\mu }|\Delta \vec{t},\lambda _{\Delta }\right>}^{2}$ is utilized in one's theoretical or experimental model.  That is because
$\sum_{\lambda _{N}\lambda _{\Delta }}\abs{\left< N\vec{s},\lambda _{N}|j^{\mu }|\Delta \vec{t},\lambda _{\Delta }\right>}^{2}=\sum_{\lambda _{N}\lambda _{\Delta }}\abs{<N\overrightarrow{{\Lambda_{\chi} s}}{\lambda _{N}}|\Lambda_{\chi}j^{\mu }{\Lambda_{\chi}^{-1}}|\Delta \vec{0}{\lambda _{\Delta }}>}^{2}$ since $\sum_{m''}D_{{m''}\;m}^{(j)*}(R)D_{{m''}\;m'}^{(j)}(R)=\delta_{m\;m'}$\cite{rose1957:etam}.





\bibliographystyle{apsrev}

\end{document}